\documentclass[12pt]{iopart}
\usepackage{iopams}  
\usepackage{graphicx}
\begin{document}

\title
[Guided modes in photonic structures with left-handed components]
{Guided modes in photonic structures\\ with left-handed components}

\author{P Marko\v{s}}

\address{Faculty of Mathematics, Physics and Informatics, Comenius University in Bratislava
  Mlynsk\'a dolina 2, 842 28 Bratislava, Slovakia} 
\eads{\mailto{peter.markos@fmph.uniba.sk}}

\begin{abstract}
The spectrum of guided modes of  
linear chain of dielectric and left-handed cylinders is analyzed. The structure of 
eigenfrequences is much more richer  if cylinders are made from the left-handed material 
with both permittivity
and permeability negative. The number of guided modes is much larger, and their interaction with incident
electromagnetic wave is much stronger. 
For some value of the wave vector, no guided modes were found. 
We discuss how these 
 specific properties of guided modes correspond to folded bands, observed recently 
in  photonic structures
with left-handed components.  \end{abstract}

\vspace{2pc}

\submitto{\JPB}
\maketitle

\def\calh{{\cal H}}
\def\calhw{{\cal H}(w)}
\def\calhn{{\cal H}(un)}
\def\cali{{\cal J}}
\def\phi{\varphi}

\section{Introduction}

Spatial periodicity of photonic structures composed from two dielectric materials  is responsible for a broad variety of interesting physical phenomena. 
Band  structure of frequency spectrum,  appearance of gaps of forbidden frequencies
\cite{joan-pc,sakoda,s-pc}, bound states \cite{pc-asym}, surface states \cite{pc-spp} and guided modes in periodic structures 
\cite{fan,sfan}
represent only a few of them.
 Artificial metal  -- composite  with negative effective permittivity
can be constructed when one material is supplied by metal \cite{pendry-j,efros,pm-ol}.

Discovery of left-handed materials
\cite{science,ez}   with simultaneously negative permittivity and permeability, naturally addresses
the question whether such materials could be used in the construction of new photonic structures
and how negativity's of permittivity and permeability influence properties of resulting composite.
Already in one dimensional systems, a new gap associated with zero mean value of refractive index was 
observed 
\cite{li}.
In two dimensional photonic crystals, the use of left-handed components led to the 
existence of folded bands 
with zero density of states for a certain interval of wave vectors 
\cite{busch,chen}.  
Numerical calculation of the transmission coefficients for photonic crystals with left-handed cylinders 
led to unexpected numerical instabilities  when standard numerical methods were used
\cite{kajtar}. Typical example of such instability is shown  below in figure \ref{fig:lhm-t} which shows
numerical data for  transmission coefficient obtained by  the transfer matrix algorithm for
different spatial discretization  give inconsistent results, which do not converge even in the limit of
the most  accurate accessible numerical approximation. Physically, such numerical instability indicates that
the spatial distribution of electromagnetic field inside the photonic structure is strongly inhomogeneous
\cite{aaa}.

In  this paper we suggest that unusual properties of periodic  photonic structures with left-handed components are due to
excitation of broad spectra of guided modes which interact with incident electromagnetic wave.
We show that indeed  photonic structures with left-handed component support the excitation of a broad variety of spatially 
inhomogeneous  guided modes which strongly influence the  transmission of electromagnetic waves. 

The paper is organized as follows:
in Section \ref{sect:model} we  describe  the model and derive mathematical equations for guided modes in  
the most simple photonic structure: the linear chain of cylinders.
 Obtained system of linear equations 
is used  in Section \ref{sect:spectrum}  to calculate of  complete spectrum of guided modes both inside  and outside the light cone.
In Section \ref{sect:transm}
we use  our numerical  method for the calculation of the transmission through of plane electromagnetic wave through our structures. 
For dielectric cylinders, we proved  that Fano resonance observed in the transmission spectra are due to the excitation 
of leaky guided modes \cite{joan-pc,fan,sfan}. 
Much stronger interaction of transmitted wave  with guided modes is found in
structures  containing left-handed cylinders. 
Conclusion is given in Section \ref{sect:concl}.

\section{The model}\label{sect:model}

\begin{figure}[t]
\begin{center}
\includegraphics[width=0.4\textwidth]{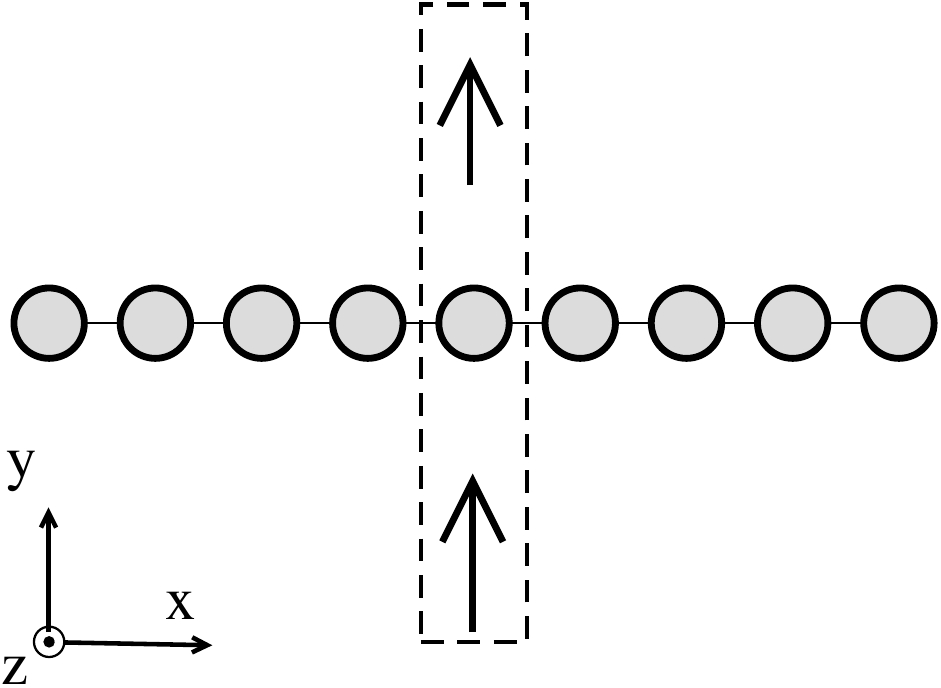}
\caption{Structure consists from periodic linear chain of homogeneous cylinders of infinite length in the $z$ direction.
Cylinders have relative  permittivity $\varepsilon$ and permeability $\mu$. We consider $\varepsilon = 12$ and $\mu=1$
for the dielectric cylinder, and $\varepsilon = -12$ and $\mu=-1$ when cylinders are made from the left-handed material.
The chain is located along the line $y=0$. Central cylinder is located at $x,y = 0,0$.  
The structure is periodic in the $x$ direction with period $a$. 
The radius of cylinders is $R=0.3 a$.
Electromagnetic  wave impinges on the structure  in $y$ direction  with polarization $E\parallel z$ or $H\parallel z$. 
Since the structure is periodic in the $x$ direction, it is sufficient to consider only one unit cell. 
}
\label{fig:geom}
\end{center}
\end{figure}

The structure of interest consists of an infinite periodic  chain of cylinders
embedded in the vacuum.
 (figure \ref{fig:geom})
Distance between cylinders, $a$, defines spatial periodicity of the structure along the $x$ direction. 
Cylinders are infinite along the $z$ direction, have radius $R=0.3a$ 
and are made from homogeneous material with 
relative permittivity $\varepsilon$  and permeability $\mu$. We chose 
$\varepsilon=+12$, $\mu=+1$ for dielectric cylinders, and
$\varepsilon=-12$, $\mu=-1$ for  cylinders made from left-handed material. 

We are interested in eigenstates propagating along  the $x$ direction with wave vector $q$
and frequency $\omega$. If $(q,\omega)$ lies inside the light cone (leaky modes with $q<\omega/c$), these states
could couple to  incident electromagnetic field.

\subsection{Equations for guided modes}


Consider first the $E_z$ polarized electromagnetic wave. For the central cylinder, we express 
the electric intensity $e_z$ and tangential component of the magnetic intensity $h_t$   in terms of Bessel functions
\cite{stratton}. Inside the cylinder ($r\le R$) we find
\begin{equation}
\label{eq:inc}
\begin{array}{rclcl}
e_z^{\rm in}(r,\phi) &=& {\cali}_0\alpha_0^+ 
&+& 2\displaystyle{\sum_{k>0} }\alpha_k^+{\cali}_k \cos(k\phi) + 
 2i\displaystyle{\sum_{k>0}} \alpha_k^-{\cali}_k \sin(k\phi)\\
&&&& \\
-iZh_t^{\rm in}(r,\phi) &=& {\cali}'_0\alpha_0^+ 
&+& 2\displaystyle{\sum_{k>0}} \alpha_k^+{\cali}'_k \cos(k\phi) +  
2i\displaystyle{\sum_{k>0}} \alpha_k^-{\cali}'_k \sin(k\phi)
\end{array}
\end{equation}
In equation  \ref{eq:inc}  we used 
\begin{equation}
{\cali}_k = J_k(2\pi n r/\lambda),~~~
{\cali}'_k = J'_k(2\pi n r/\lambda)
\end{equation}
where  $J_k$, $J'_k$  are  the Bessel function ($k=0,1,\dots$) and their derivatives,
$\lambda=2\pi c/\omega$ is the wavelength of electromagnetic field in vacuum,
$n=\sqrt{\varepsilon\mu}$ is the 
index of refraction of the cylinder and $Z=Z_0\sqrt{\mu/\varepsilon}$ ($c$ is the light velocity in vacuum and $Z_0$ is the vacuum impedance).
Unknown parameters $\alpha^+$ and $\alpha^-$ determines amplitudes of even ($\propto \cos k\phi$) and odd ($\propto\sin k\phi$)
cylindrical waves, respectively.

\smallskip

The electric and magnetic fields outside the central cylinder  consist from  two contributions: the first one is the field  scattered 
on  the central cylinder,
\begin{equation}
\label{eq:onc}
\begin{array}{rclcl}
e_z^0(r,\phi) &=& {\calh}_0\beta_0^+ 
&+& 2\displaystyle{\sum_{k>0}} \beta_k^+{\calh}_k \cos(k\phi)
+  2i\displaystyle{\sum_{k>0}} \beta_k^-{\calh}_k \sin(k\phi)\\
&&&& \\
-i Z_0h_t^0(r,\phi) &=& {\calh}'_0\beta_0^+ 
&+& 2\displaystyle{\sum_{k>0}} \beta_k^+{\calh}'_k \cos(k\phi)   
+ 2i\displaystyle{\sum_{k>0}} \beta_k^-{\calh}'_k \sin(k\phi)
\end{array}
\end{equation}
($r\ge R$)
with free parameters $\beta^+$ and $\beta^-$. Here,
\begin{equation}
\label{eq:haha}
{\calh}_k = H_k(2\pi r/\lambda),~~~
{\calh}'_k = H'_k(2\pi r/\lambda).
\end{equation}
and 
$H_k(z) = J_k(z) + i Y_k(z) $ is the first Hankel function
\cite{stratton,AS}.

\begin{figure}[t]
\begin{center}
\includegraphics[width=0.45\textwidth]{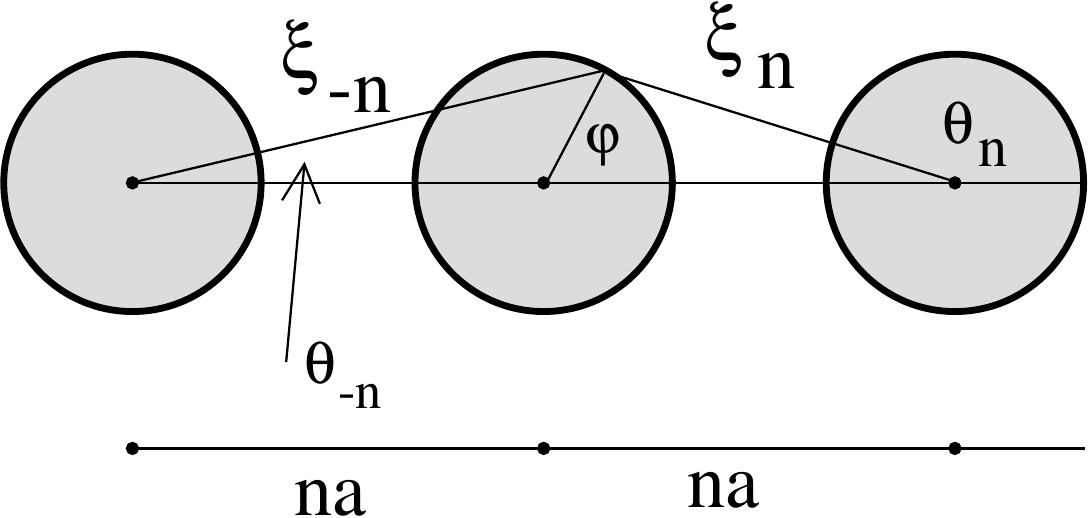}
\caption{Parameters used in derivation of equation \ref{eq:m}.
}
\label{fig:gegen_n}
\end{center}
\end{figure}

The second contribution to the external fields is represented by fields scattered from all other cylinders. 
For the  $n$th one  ($n=\pm 1, \pm 2,\dots ,N_s$)
we express electric and magnetic intensity  in coordinates associated with 
the center of the  cylinder (figure \ref{fig:gegen_n})
\begin{equation}\label{eq:m}
\begin{array}{rclcl}
e_z^n(\xi_n,\theta_n) &=& H_0(w_n)\beta_{n0}^{+} 
&+& 2\displaystyle{\sum_{k>0}} \beta_{nk}^+H_k(w_n) \cos(k\theta_n)\\
&& &+& 2i\displaystyle{\sum_{k>0}} \beta_{nk}^-H_k(w_n) \sin(k\theta_n)\\
 &&&& \\
-iZ_0h_{t_n}^n(\xi_n,\theta_n) &=& H'_0(w_n)\beta_{n0}^+ 
&+& 2\displaystyle{\sum_{k>0}} \beta_{nk}^+H'_k(w_n) \cos(k\theta_n)   \\
&& &+& 2i\displaystyle{\sum_{k>0}} \beta_{nk}^-H'_k(w_n) \sin(k\theta_n)\\
 &&&& \\
Z_0h_{r_n}^n(\xi_n,\theta_n) &=& H_0(w_n)\beta_{n0}^+ 
&+& 2\displaystyle{\sum_{k>0}} \beta_{nk}^+\frac{k}{w_n}H_k(w_n) \cos(k\theta_n)   \\
&& &+& 2i\displaystyle{\sum_{k>0}} \beta_{nk}^-\frac{k}{w_n}H_k(w_n) \sin(k\theta_n)
\end{array}
\end{equation}
where we use
\begin{equation}
 w_n = 2\pi \xi_n/\lambda
\end{equation}

Thanks to the periodicity of the model, coefficients $\beta_{nk}^\pm$ of the $n$th cylinder could be 
expressed in terms of coefficients $\beta_k^\pm$ as 
\begin{equation}
\beta_{nk}^\pm = e^{iqna}\beta_k^\pm
\end{equation}

Coefficients $\alpha$ and $\beta$ could be calculated from the requirement of the  
continuity of fields $e_z$ and $h_\phi$ at the surface of the cylinder:
\begin{equation}
\label{eq:lin-e}
e_z^{\rm in}(R^-,\phi) = e_z^{\rm out}(R^+,\phi) = e_z^0(R^+) + \sum_{n\ne 0} e_z^n(\zeta_n,\theta_n)
\end{equation}
and
\begin{equation}
\label{eq:lin-h}
h_t^{\rm in}(R^-) = h_t^{\rm out} = h_t^0(R^+,\phi) + \sum_{n\ne 0} \left[ h_{t_n}^n(\xi_n,\theta_n)\cos \alpha_n  - h^n_{r_n}(w_n,\theta_n)\sin \alpha_n \right]
\end{equation}
where $\alpha_n = \theta_n-\phi$.

\begin{figure}[b]
\begin{center}
\includegraphics[width=0.35\textwidth]{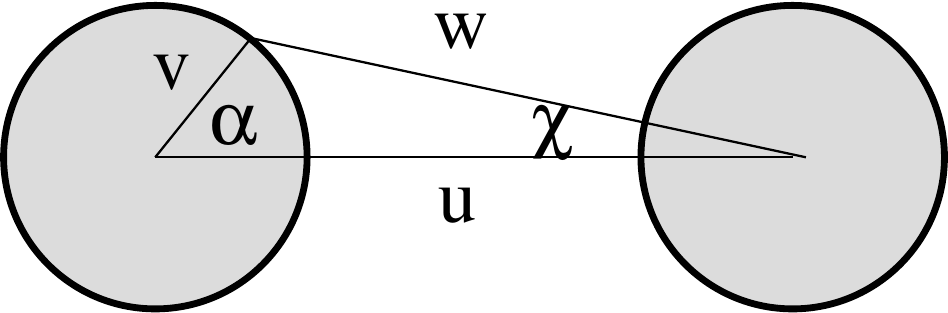}
\caption{Parameters  used in  Gegenbauer's relation, equations \ref{eq:geg} and  \ref{eq:gegen}
}
\label{fig:gegen}
\end{center}
\end{figure}

Before  solving  this system of equations, we  first transform all fields in equation \ref{eq:m} 
as a functions of  $R$ and $\phi$. This can be done 
with the use of the Gegenbauer formula for cylindrical functions (figure  \ref{fig:gegen})
\begin{equation}\label{eq:geg}
H_m(w) e^{\pm im\chi} = \sum_{k=-\infty}^{+\infty}  H_{m+k}(u)J_k(v)e^{\pm ik\alpha}
\end{equation}
\cite{AS} 
and similar formula for the first derivative $H'_m(w)$: 
\begin{equation}
\label{eq:gegen}
H'_m(w) e^{\pm im\chi} = \sum_{k=-\infty}^{+\infty}  H'_{m+k}(u)J_k(v)e^{\pm ik\alpha}
\end{equation}
Inserting (\ref{eq:gegen}) into  equations (\ref{eq:m}) 
we express, after some algebra,  the fields at the outer boundary of the cylinder in the form
\begin{equation}
\label{eq:continuity}
\begin{array}{rclcl}
e_z^{\rm out}(R^+,\phi) &=& 
\displaystyle{\sum_{k,m=0}^{N}} \textbf{B}_{km}\beta^+_m\cos k\phi &+& 
\displaystyle{\sum_{k,m=1}^N}\textbf{C}_{km}\beta^-_m\sin k\phi\\
h_t^{\rm out}(R^+,\phi) &=&
\displaystyle{\sum_{k,m=0}^{N}} \textbf{B'}_{km}\beta^+_m\cos k\phi &+& 
\displaystyle{\sum_{k,m=1}^N} \textbf{C'}_{km}\beta^-_m\sin k\phi
\end{array}
\end{equation}
Where we introduced $N$ as the highest order Bessel function used in the expressions of the fields.
Explicit form of matrices
$\textbf{B}$,
$\textbf{B'}$,
$\textbf{C}$ and
$\textbf{C'}$
is given in the Appendix A.

Incident electromagnetic field  contributes  to the rhs of equation \ref{eq:continuity}. 
We express it in the cylindrical coordinates
\begin{equation}
\label{eq:ini}
\begin{array}{rclclcl}
e^i_z(r,\phi) &=& J_0(v) 
&+& \displaystyle{\sum_{k>0}^N} e_k^+ \cos(k\phi)  
 &+& i \displaystyle{\sum_{k>0}^N} e_k^- \sin(k\phi)\\
-iZ_0h^i_\phi(r,\phi) &=& J'_0(v) 
&+& \displaystyle{\sum_{k>0}^N} h_k^+ \cos(k\phi)   
 &+& i \displaystyle{\sum_{k>0}^N} h_k^- \sin(k\phi)
\end{array}
\end{equation}

Inserting fields (\ref{eq:inc}) and  \ref{eq:continuity}) 
into equations \ref{eq:lin-e} and \ref{eq:lin-h} we obtain  
the set of linear equations for even modes:
\begin{equation}
\label{eq:lin1}
\begin{array}{rcl}
\textbf{A}_k \alpha^+_k = \sum_m\textbf{B}_{km}\beta^+_m + e_k^+\\
&& \\
\zeta \textbf{A'}_k\alpha^+_k = \sum_m\textbf{B'}_{km} \beta^+_m + h_k^+
\end{array}
\end{equation}
($k=0,1,\dots, N$)
and similar equations for odd modes
\begin{equation}
\label{eq:lin1a}
\begin{array}{rcl}
\textbf{A}_k \alpha^-_k = \sum_m\textbf{C}_{km}\beta^-_m + e_k^-\\
&& \\
\zeta \textbf{A'}_k\alpha^-_k = \sum_m\textbf{C'}_{km} \beta^-_m + h_k^-
\end{array}
\end{equation}
($k=1,2,\dots, N$).
Here,  $\zeta = Z/Z_0$, 
$\textbf{A}_{k} = {\cali}_k (2-\delta_{k0})$
and $\textbf{A'}_{k} = {\cali}'_k (2-\delta_{k0})$
(equation \ref{eq:inc}).

In the next step,
we remove $\alpha^+$ from equations \ref{eq:lin1} and  obtain the system of linear
equations for parameters $\beta^+$:
\begin{equation}
\label{eq:lin2}
\sum_m \left[\textbf{B}_{km} - \zeta\frac{{\cali}_k}{{\cali}'_k}\textbf{B'}_{km}\right]\beta^+_m = 
e_k^+ - \zeta\frac{{\cali}_k}{{\cali}'_k} h_k^+ 
,~~~~~~~k = 0,1,\dots ,N 
\end{equation}
and  equivalent equation for parameters $\beta^-$:
\begin{equation}
\label{eq:lin3}
\sum_m \left[\textbf{C}_{km} - \zeta\frac{{\cali}_k}{{\cali}'_k}\textbf{C'}_{km}\right]\beta^-_m = 
e_k^- - \zeta\frac{{\cali}_k}{{\cali}'_k} h_k^-,~~~~~~k = 1,2,\dots, N 
\end{equation}

With known parameters $\beta^\pm$, we find $\alpha^\pm$ from  equations \ref{eq:lin1} and \ref{eq:lin2}.

In numerical analysis,
we consider the number of cylinders $N_s=100 - 4000$ and number of modes $N\le 22$. We found that consideration of  $N=12$
modes is sufficient in most of analyzed problems.

\subsection{Transmission coefficient}

Transmission coefficient can be calculated as the ratio of the $y$-component of the Poynting vector $S_x(y_p)$,
calculated for any $y_p>r$
to the incident Poynting vector, $S_y^i$
\begin{equation}
\label{eq:T}
T = \frac{S_x(y_p)}{S_x^i} = 
\int_{-a/2}^{+a/2} {\rm d} x~ e_z(x,y_p) h^*_x(x,y_p)  ~/~
\int_{-a/2}^{+a/2} {\rm d} x~ e^i_z(x,y_p) (h^i_x(x,y_p) )^*
\end{equation}

\subsection{Polarization}

The above analysis Can be repeated for the $H_z$ polarized modes. 
It turns out that final equations (\ref{eq:lin2}) and {\ref{eq:lin3}) remain the same if we substitute 
\begin{equation}
\zeta \to \frac{1}{\zeta}
\end{equation}
and exchange of $e_x$ and $h_z$ in expression for the transmission coefficient (\ref{eq:T}).


\section{Guided modes}\label{sect:spectrum}

\begin{figure}[t!]
\begin{center}
\includegraphics[width=0.35\textwidth]{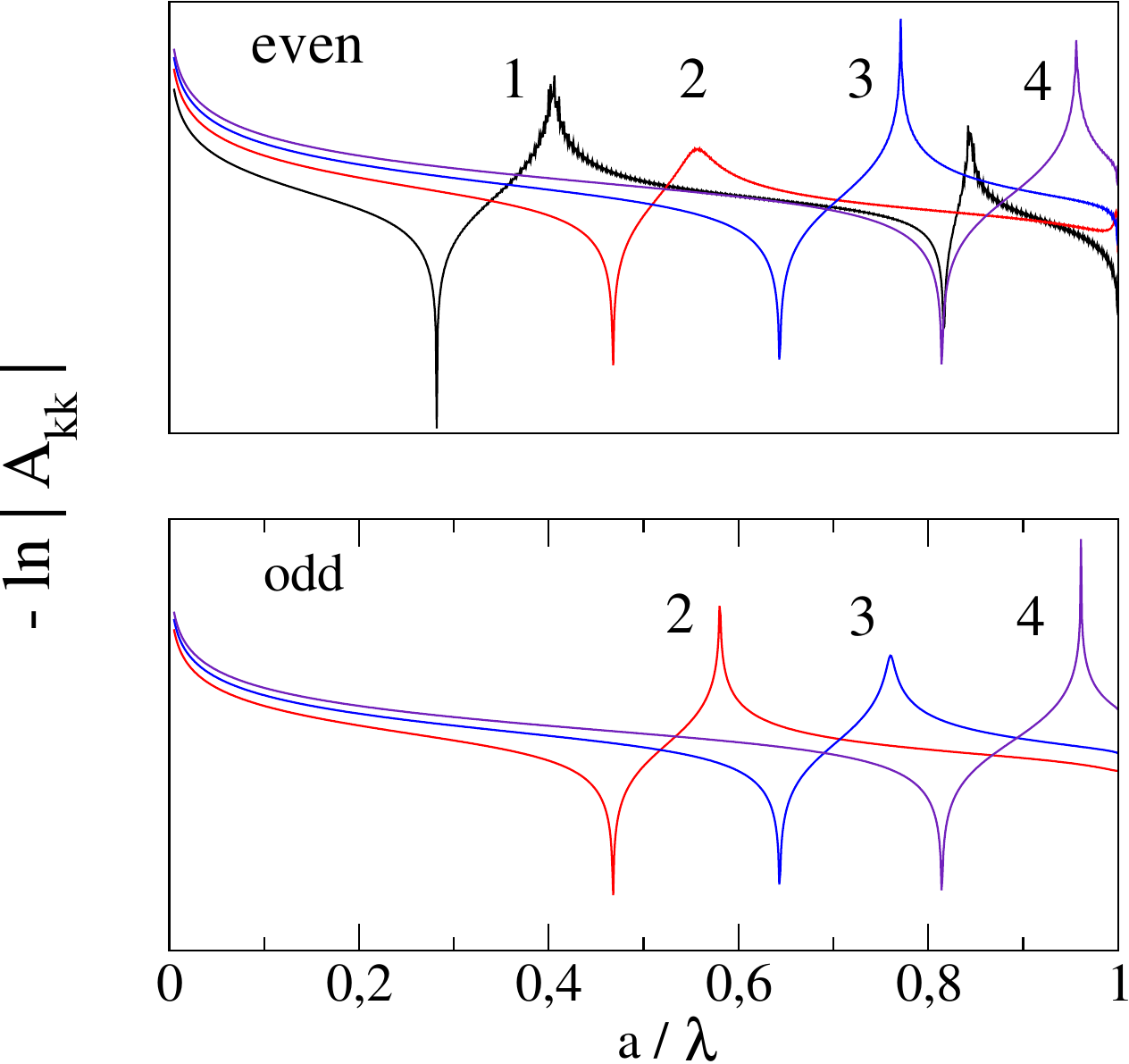}
~~~~~~
\includegraphics[width=0.35\textwidth]{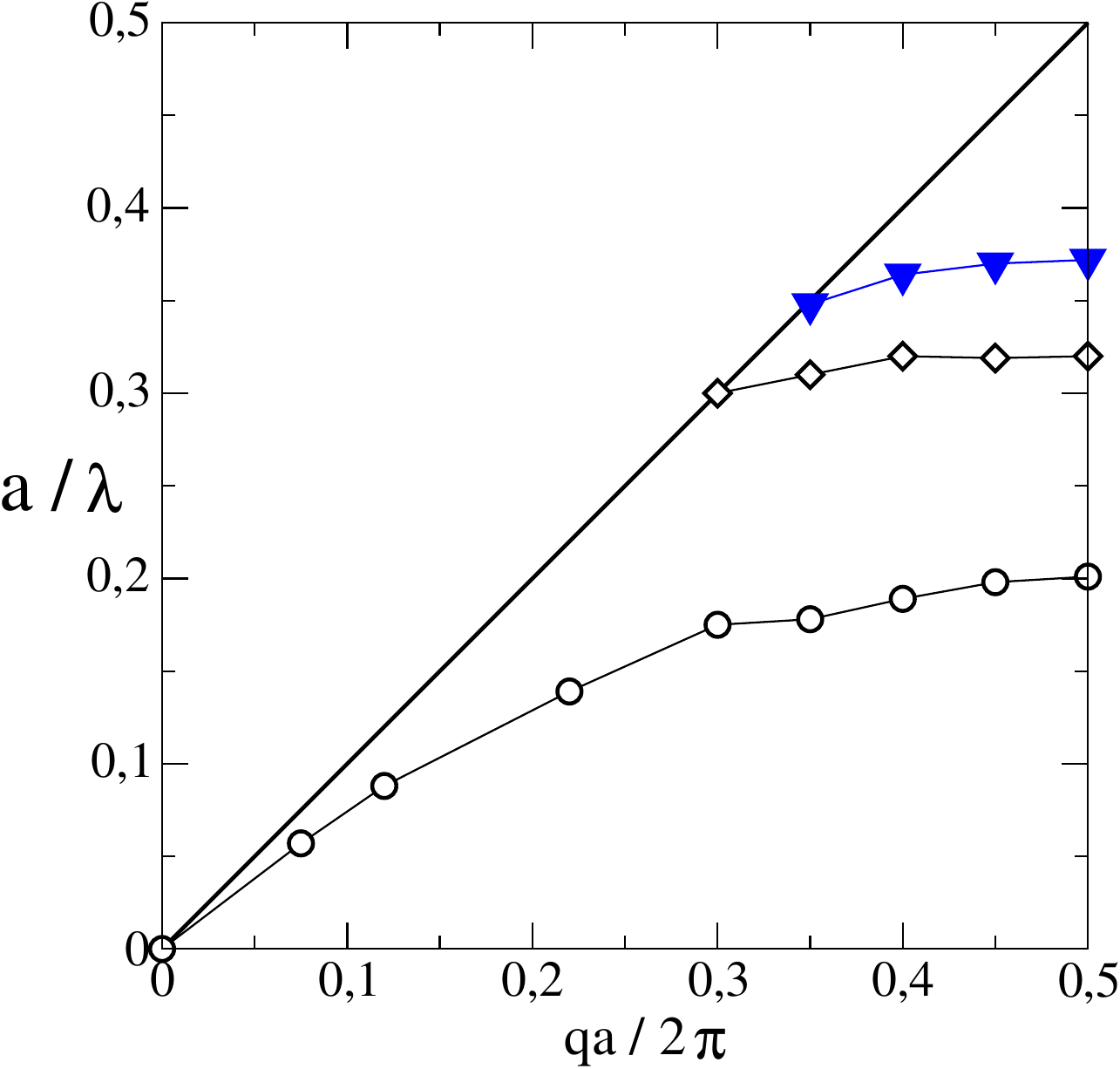}
\caption{Guided modes of  the  chain of dielectric cylinders ($\varepsilon=12$, $\mu=1$).
Left figure shows the frequency dependence of diagonal elements ln~$A_{kk}$ for $q=0$  (equations \ref{eq:det:even} and \ref{eq:det:odd}.
Four even modes and three odd modes can be identified.
Right figures presents  the 
spectrum  of $E_z$ polarized 
Open and closed symbols correspond to even modes ($n=0$ and 1)   and odd mode $n=1$.
Owing to the periodicity of the problem, it is sufficient to consider wave vectors $0< q<\pi/a$ \cite{joan-pc}.
}
\end{center}
\label{fig:r}
\end{figure}

Consider first equations \ref{eq:lin2} and \ref{eq:lin3} without incident electromagnetic wave.
Then, the spectrum of guided modes can be found from the requirements of zero determinant 
\begin{equation}
\label{eq:det:even}
{\rm det} 
\left[\textbf{B}_{km} - \zeta\frac{{\cali}_k}{{\cali}'_k}\textbf{B'}_{km}\right] = 0
\end{equation}
and 
\begin{equation}
\label{eq:det:odd}
{\rm det} 
\left[\textbf{C}_{km} - \zeta\frac{{\cali}_k}{{\cali}'_k}\textbf{C'}_{km}\right] = 0
\end{equation}
for even and odd modes, respectively.

With the use of Gauss-Jordan elimination method \cite{nrcp}, we  transform 
matrices in equations (\ref{eq:det:even}) and (\ref{eq:det:odd})  to diagonal form
and plot the
frequency dependence of  inverse of obtained diagonal elements $A_{kk}$. This enables us not only to find the eigenfrequency of guided modes and their  lifetime \cite{economou}, 
but also identify their  symmetry. As an example, we show in the left  figure 4
resonances  of leaky guided modes with zero wave vector $q=0$. 

Right figure 4 presents eigenfrequences of guided modes as a function of wave vector $q$  identified 
from the position of resonances in corresponding diagonal elements.
We found two even modes with $k=0$ and 1 and one odd mode $k=1$, which  agree with our expectation \cite{joan-pc}.

\medskip

\begin{figure}[t!]
\begin{center}
\includegraphics[width=0.35\textwidth]{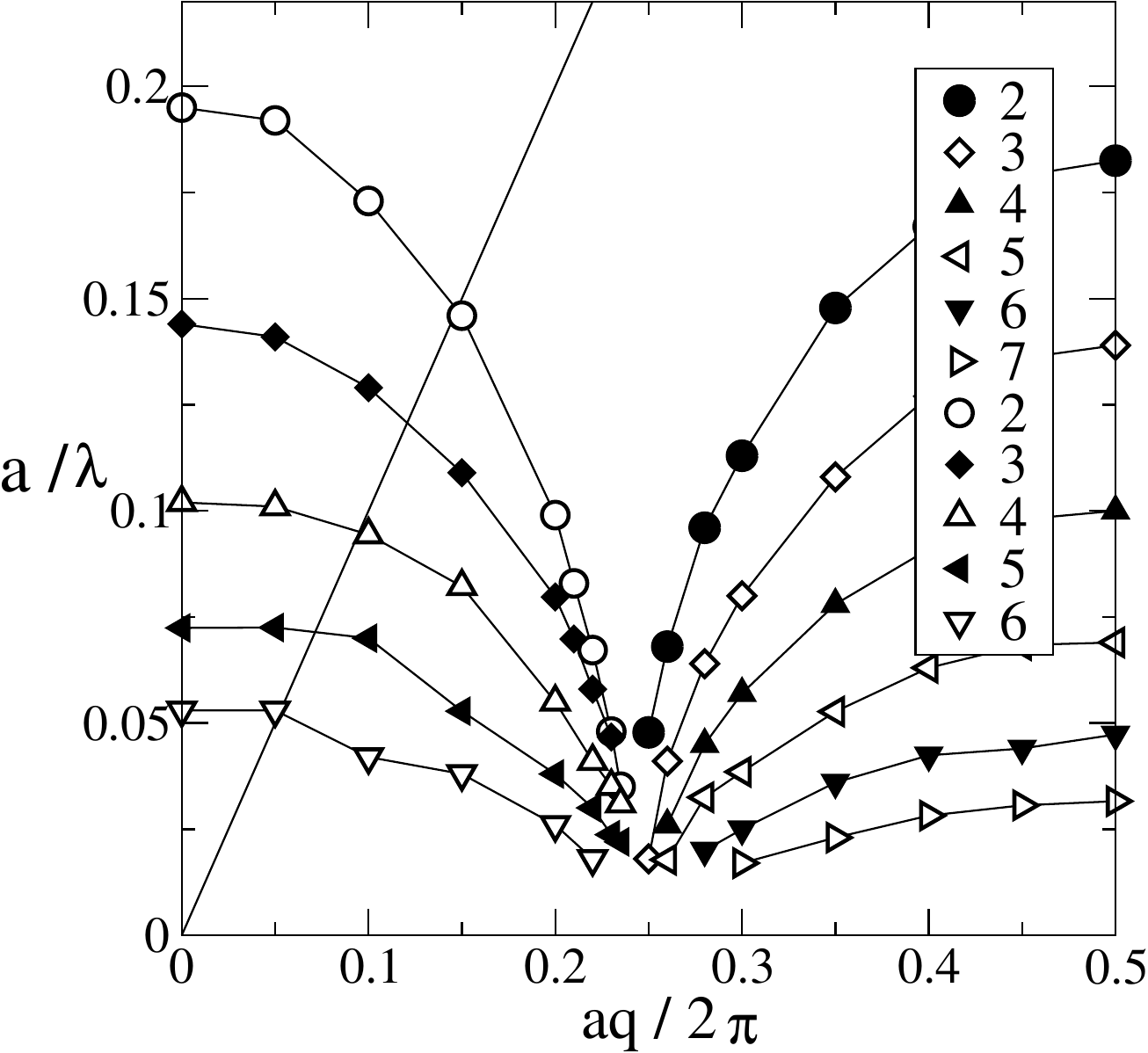}
\caption{Spectrum of guided modes of the chain of cylinders made from the  left-handed material  ($\varepsilon=-12$, $\mu=-1$)  
cylinders. There are no modes around $q_c\approx 0.24 \times 2\pi/a$.
Open and closed symbols correspond to even and odd modes, respectively. Note that guided modes changes symmetry when $q$ crosses $q_c.$}
\end{center}
\label{lhm-fig}
\end{figure}

The spectrum of guided modes for left-handed cylinders is more complicated and consists of series of even and odd modes. In contrast to dielectric cylinders, the eigenfrequences of guided modes decreases when mode index $k$ increases.
The number of modes depends on the model parameters and increases when absolute value of the refractive index increases.

Another unexpected property of guided modes is the existence of  ''critical value'' of the wave vector $q_c$
($q_c\approx 0.247 \times 2\pi/a$ in our model) for which no guided mode exists (figure 5). 
Preliminary 
analysis of  other  left-handed structures with $\mu=-1$ (data not shown)  indicates that
$q_c$ depends neither on (negative)  permittivity nor on the radius of cylinder.


\begin{figure}[t!]
\begin{center}
\includegraphics[width=0.4\textwidth]{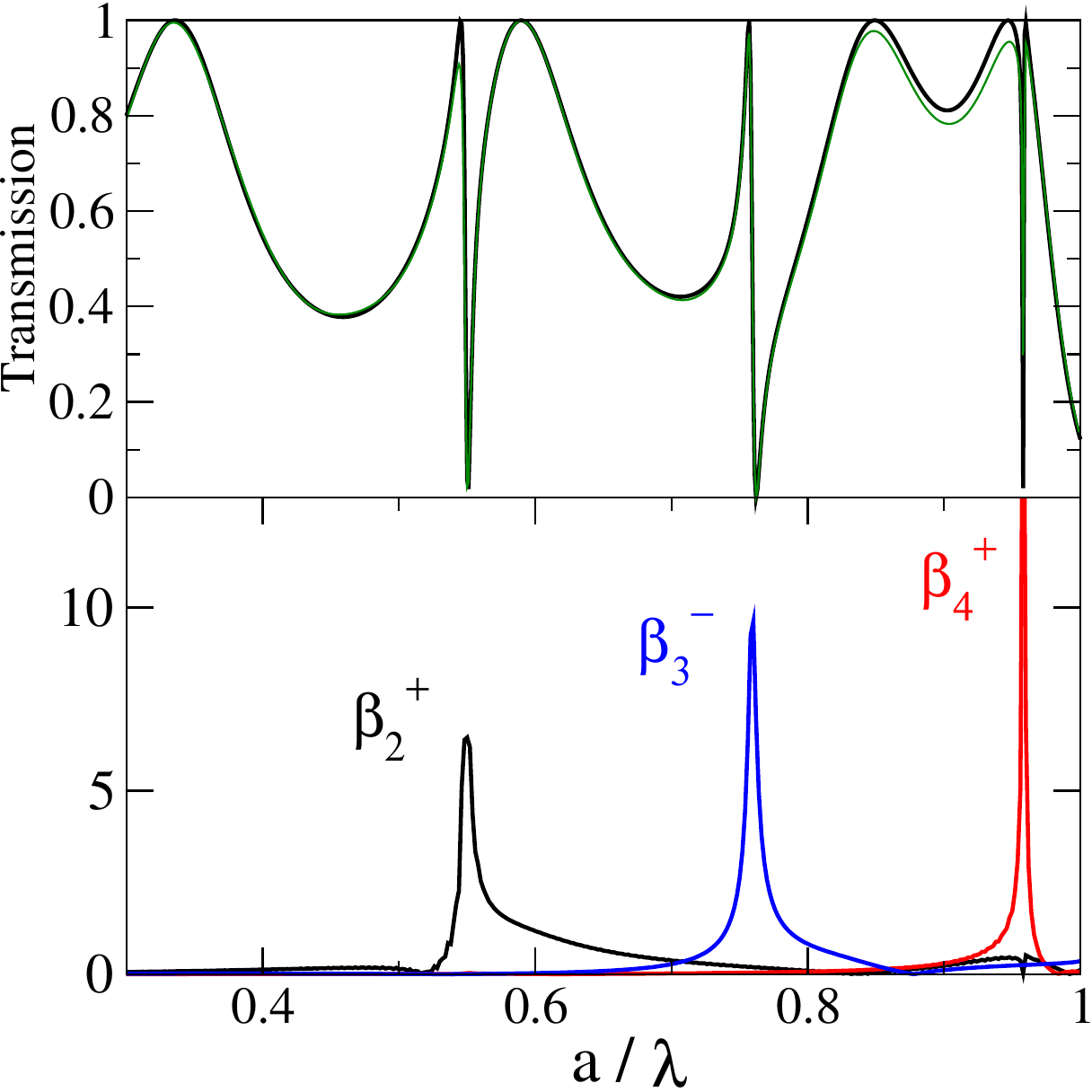}\\
\includegraphics[width=0.3\textwidth]{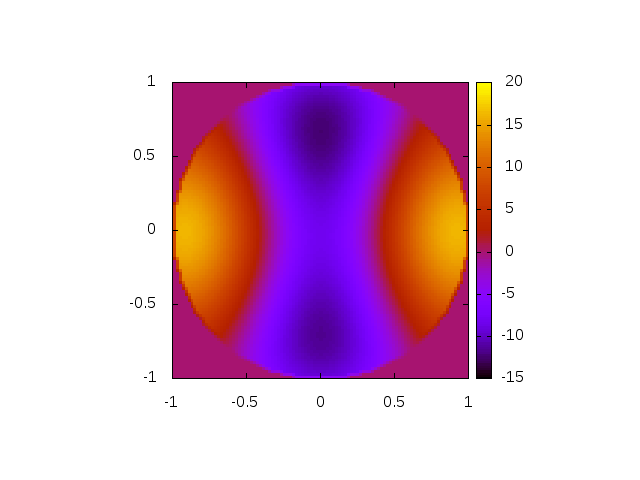}
\includegraphics[width=0.3\textwidth]{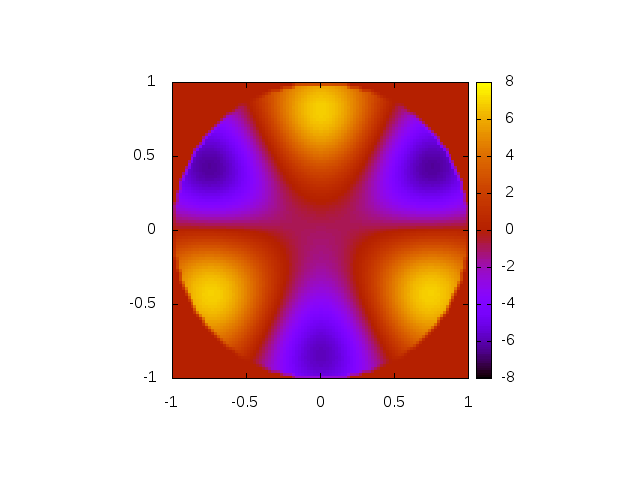}
\includegraphics[width=0.3\textwidth]{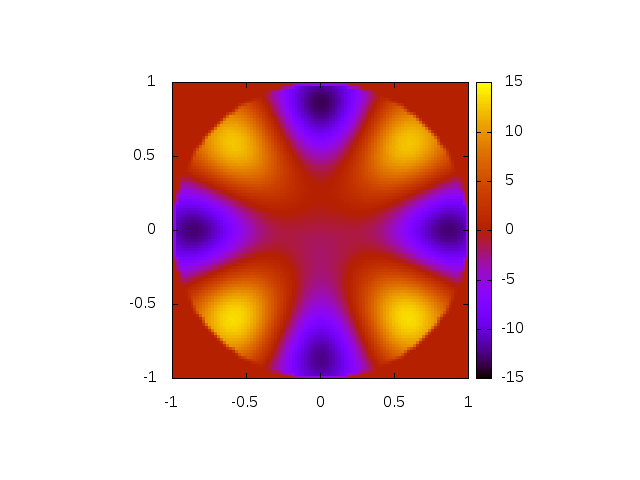}
\caption{Transmission of the $E_z$ polarized plane wave propagating through an infinite 1D chain of dielectric cylinders.
Top   panel displays 
the transmission coefficient as a function of $a/\lambda$, calculated by the transfer matrix method
and by the formula (\ref{eq:T}).
Three Fano resonances correspond
to maxims of coefficients $\beta$ shown in middle panel. 
Real part of the intensity of electric field inside the cylinder for three resonance frequencies, 
$a/\lambda = 0.549,~0.759$ and $0.958$  is shown in the bottom panel.
}
\label{fig:eps12}
\end{center}
\end{figure}

\section{Interaction of external field with guided modes}\label{sect:transm}

We solve equations \ref{eq:lin2} and \ref{eq:lin3} and calculate transmission coefficient for incident plane wave 
with wavelength $\lambda$ propagating along the $y$ direction.
First, we express the coefficients $E^\pm_k$ and $h^\pm_k$
\begin{equation}
\begin{array}{lclclr}
e_k^+  &=& h_k^+ &=& J_k(v) (1+(-1)^k)/(1+\delta_{k0})& k=0,1,\dots\\
e_k^-  &=& h_k^- &=& J_k(v) (1-(-1)^k),&k = 1,2,\dots
\end{array}
\end{equation}
($v=2\pi R/\lambda$).
 From known coefficients $\beta^+$
and $\beta^-$ we find spatial distribution of fields  and calculate transmission coefficient given by equation \ref{eq:T}.

\subsection{Dielectric cylinders}

Figure \ref{fig:eps12} shows the transmission coefficient for  the $E_z$ polarized plane electromagnetic wave 
  as a function of wavelength. 
Obtained results agree  very well  with  numerical data calculated  by the transfer matrix method 
\cite{pendry,pm}. 
Typical Fano resonances in the transmission spectra
\cite{fan,sfan}
are clearly visible and could be associated with excitation of corresponding guided modes
\cite{astr}
in cylinder chain. 
Owing to the symmetry of the incident wave which does not depend on $x$,
 only modes symmetric with respect to $x\to -x$ could be excited.
Two even modes with $k=2$ and $k=4$ and one odd mode ($k=3$) have been excited.
Bottom panels show the real part of electric field inside the cylinder and confirm predicted symmetry of excited states.

Similar results have been obtained also for the $H_z$ polarized incident wave shown in figure 7. 
In this case,  two resonances of $k=0$  mode are clearly visible. The lower one, together with the
first odd resonance, deform strongly the transmission spectrum. Three resonances observed at higher frequencies
could be easily detected both from the transmission spectra and from the  frequency dependence of parameters $\beta$.

\begin{figure}[t!]
\begin{center}
\includegraphics[width=0.4\textwidth]{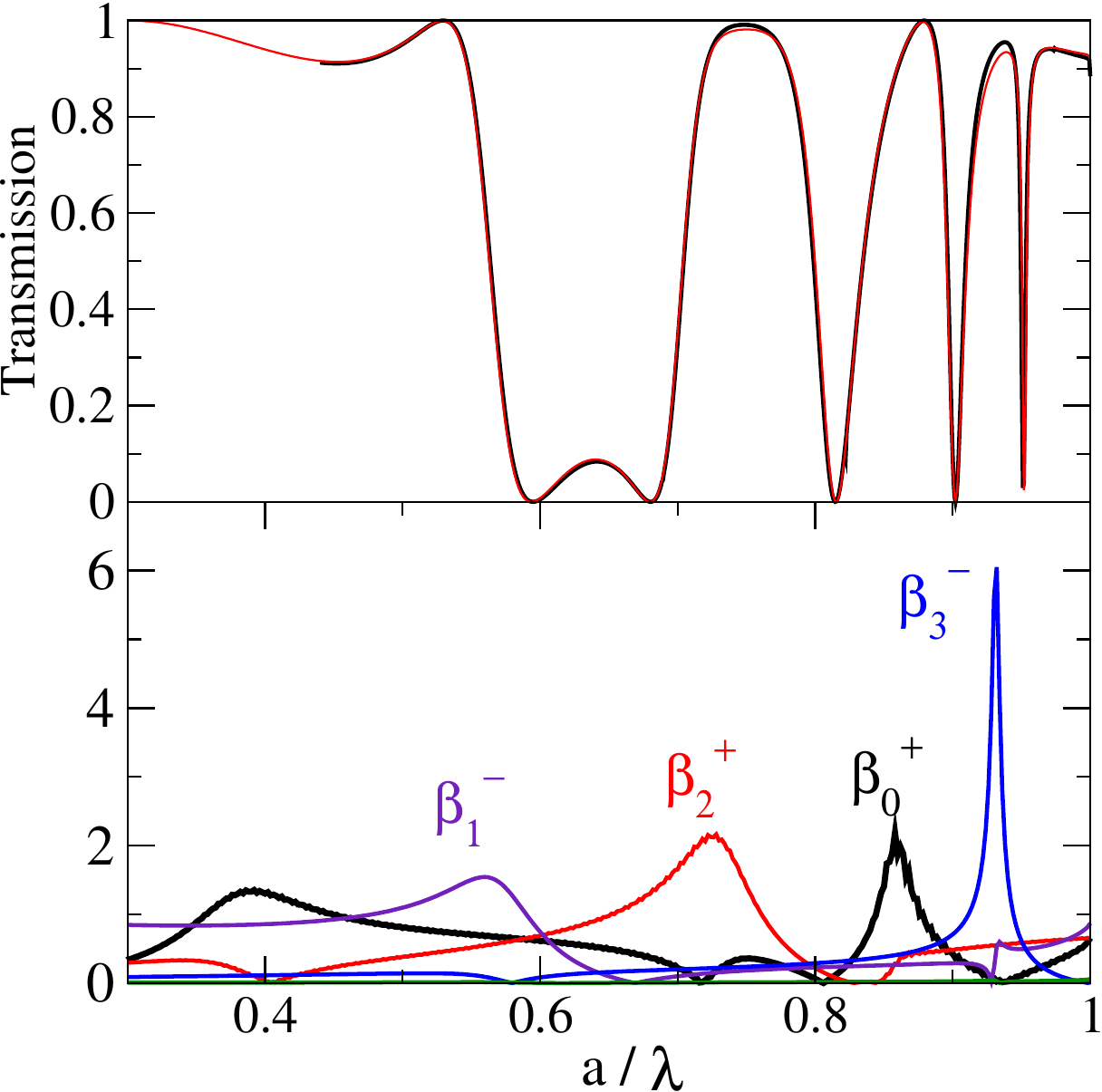}\\
\includegraphics[width=0.3\textwidth]{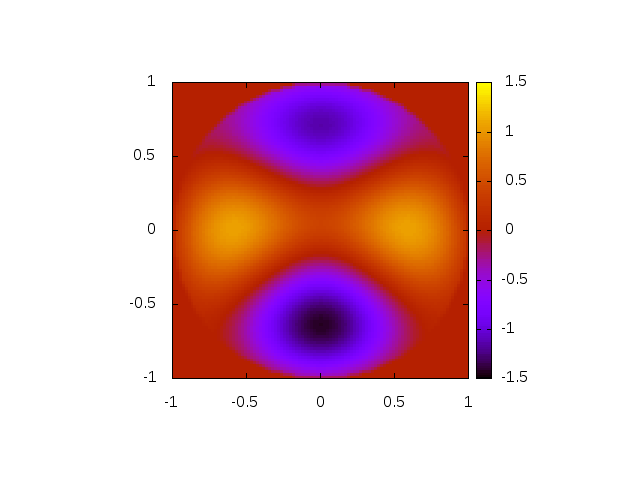}
\includegraphics[width=0.3\textwidth]{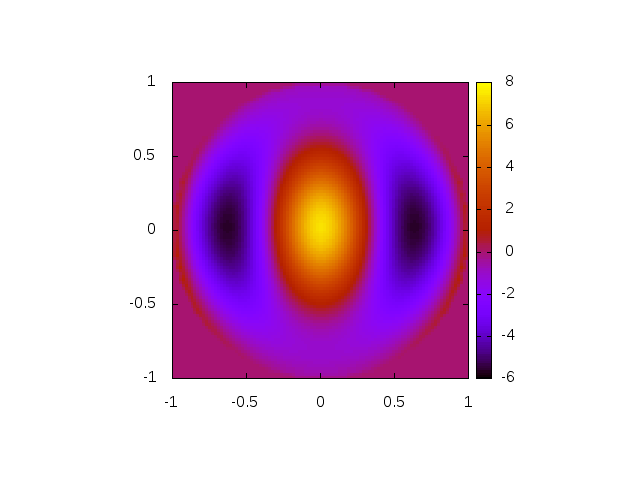}
\includegraphics[width=0.3\textwidth]{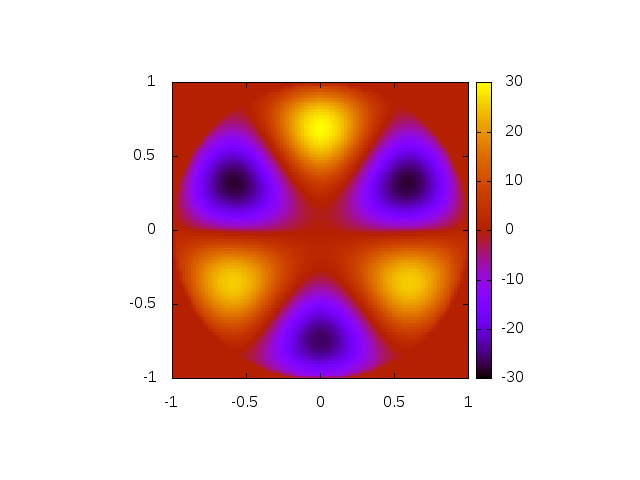}
\caption{Transmission of the $H_z$ polarized plane wave propagating through an infinite chain of dielectric cylinders.
Top   panel displays 
the transmission coefficient as a function of $a/\lambda$, calculated by the transfer matrix method
and by the formula equation \ref{eq:T}.
Fano resonances correspond
to maxims of coefficients $\beta$ shown in middle panel. 
Real part of the intensity of magnetic  field inside the cylinder for three resonance frequencies, 
$a/\lambda = 0.729,~0.865$ and $0.931$  is shown in the bottom panel.
}
\end{center}
\label{poleH}
\end{figure}

\begin{figure}[t!]
\begin{center}
\includegraphics[width=0.3\textwidth]{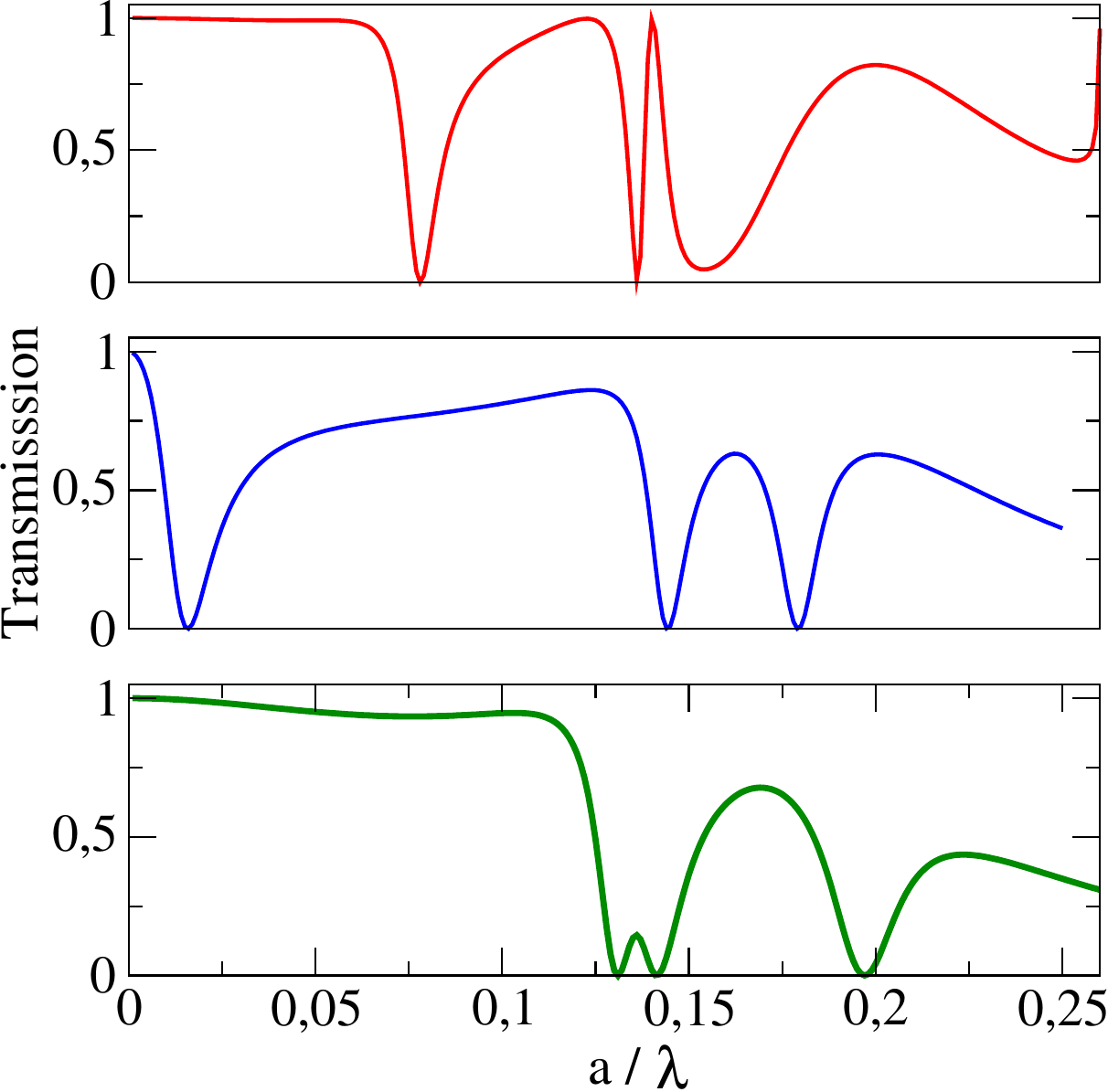}
\includegraphics[width=0.3\textwidth]{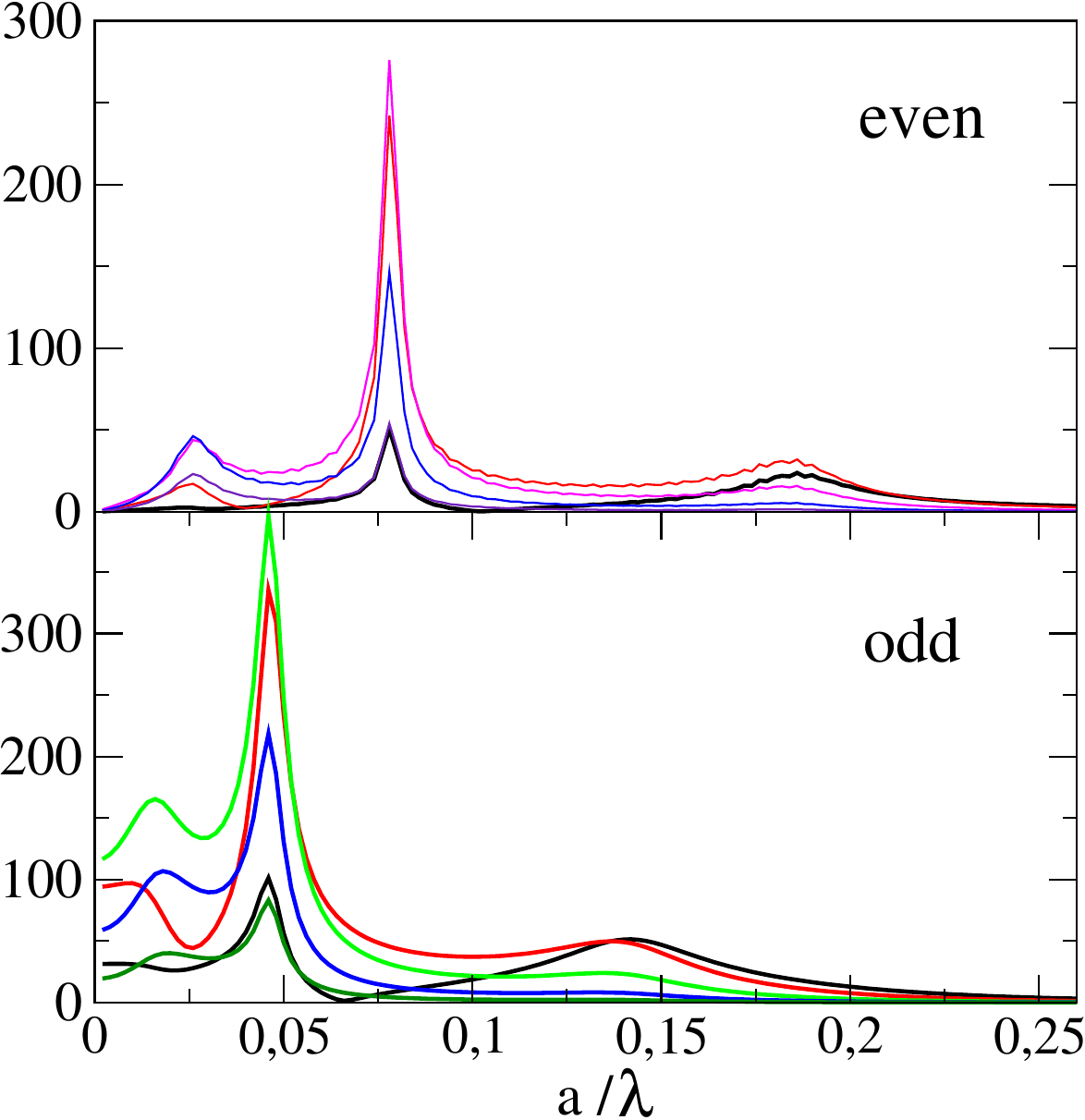}
\includegraphics[width=0.3\textwidth]{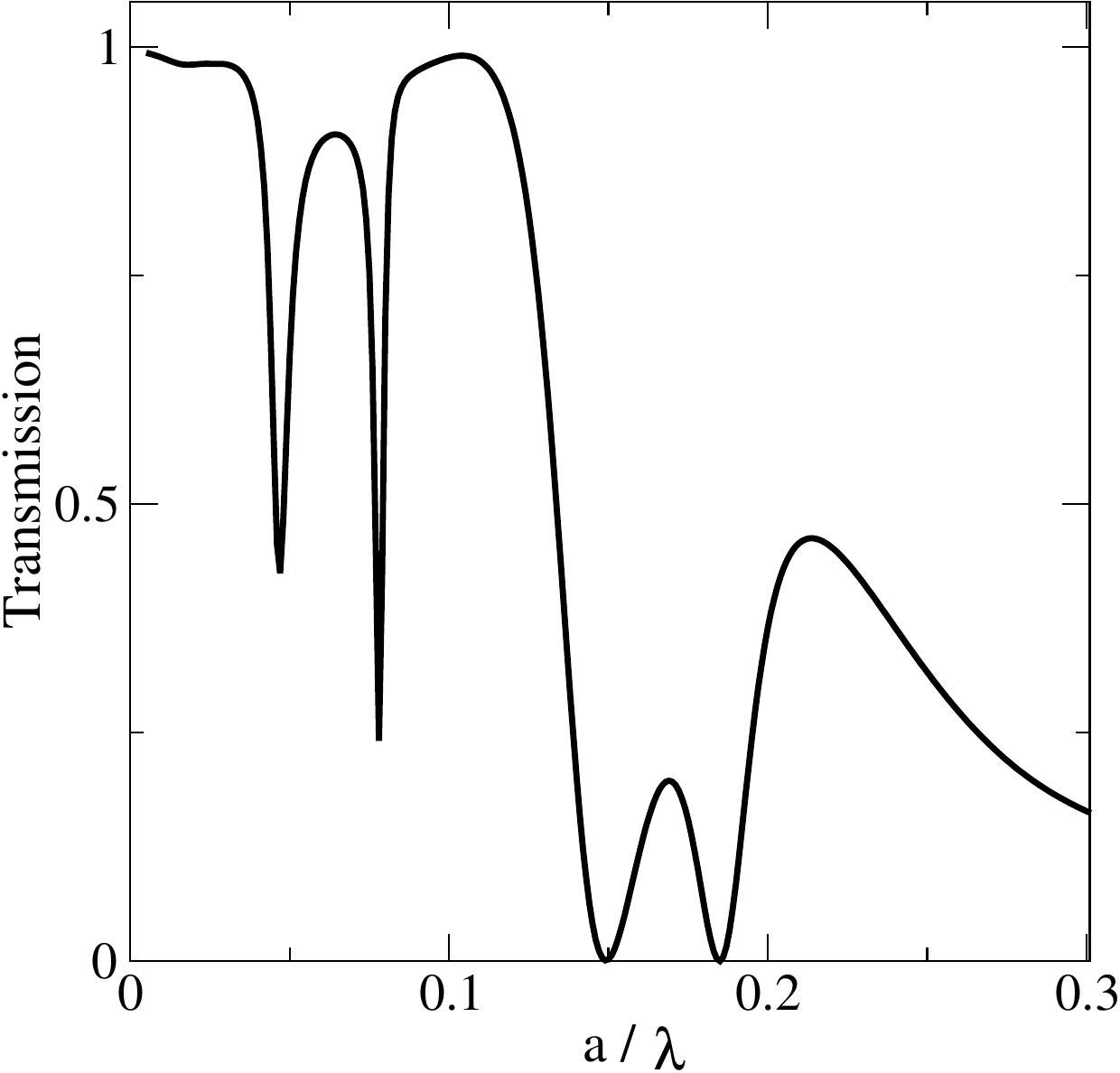}
\includegraphics[width=0.3\textwidth]{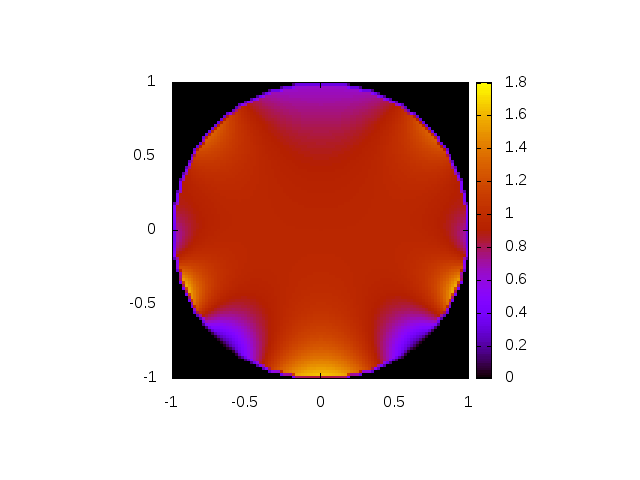}
\includegraphics[width=0.3\textwidth]{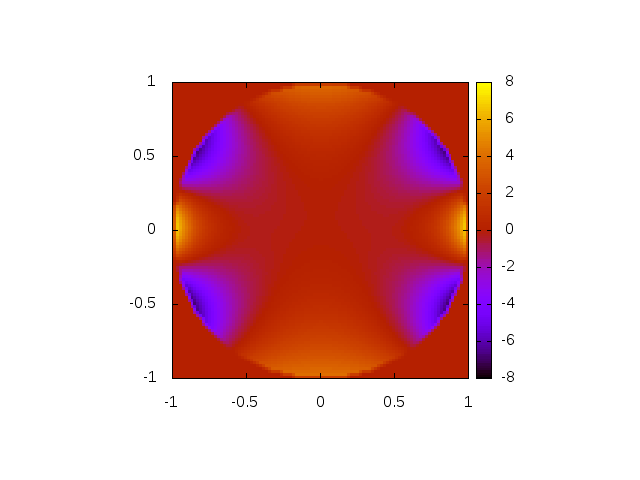}
\includegraphics[width=0.3\textwidth]{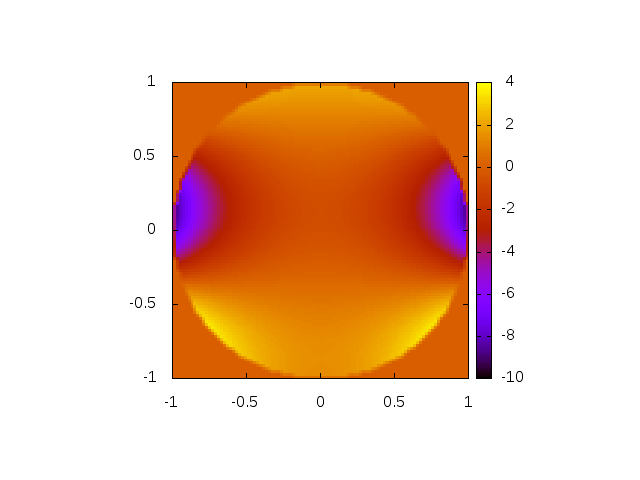}
\includegraphics[width=0.3\textwidth]{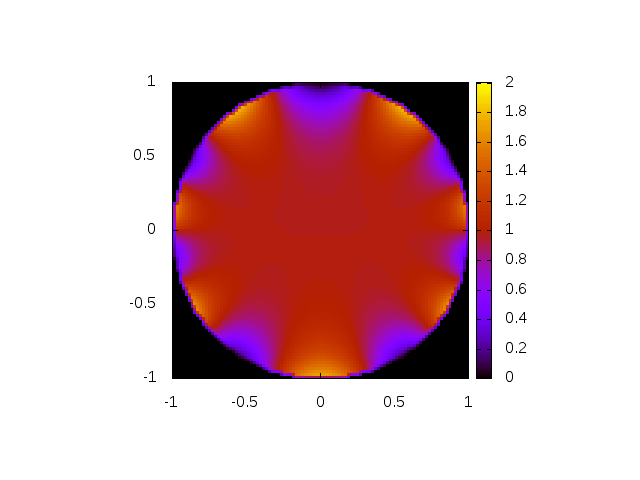}
\includegraphics[width=0.3\textwidth]{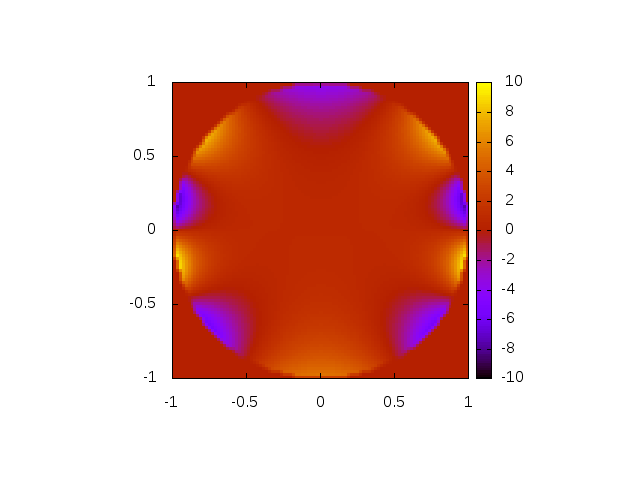}
\includegraphics[width=0.3\textwidth]{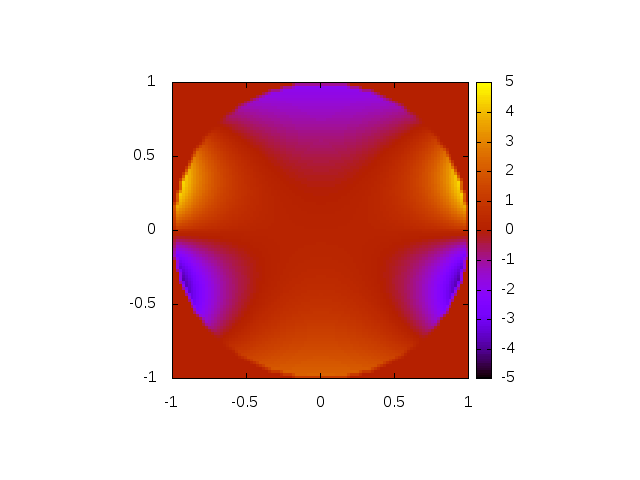}
\end{center}
\caption{Left: transmission of the $E_z$ polarized electromagnetic wave through the linear chain of left-handed cylinders 
calculated by the transfer-matrix method with three different   space discretization (120, 240 and 360 mesh
points per unit cell). 
Middle: spectrum of coefficients $\beta$. Note the scale of the vertical axis. Right panel
shows the transmission coefficient  calculated by the present method (equation \ref{eq:T} ).
Bottom panel shows real part of the intensity of electric   field inside the left-handed cylinder for frequencies associated
with resonances of even modes ($a/\lambda = 0.027,~0.078$ and $0.184$), 
and odd modes ($0.016,~0.047$ and $0.140$).}
\label{fig:lhm-t}
\end{figure}

\begin{figure}[t!]
\begin{center}
\includegraphics[width=0.35\textwidth]{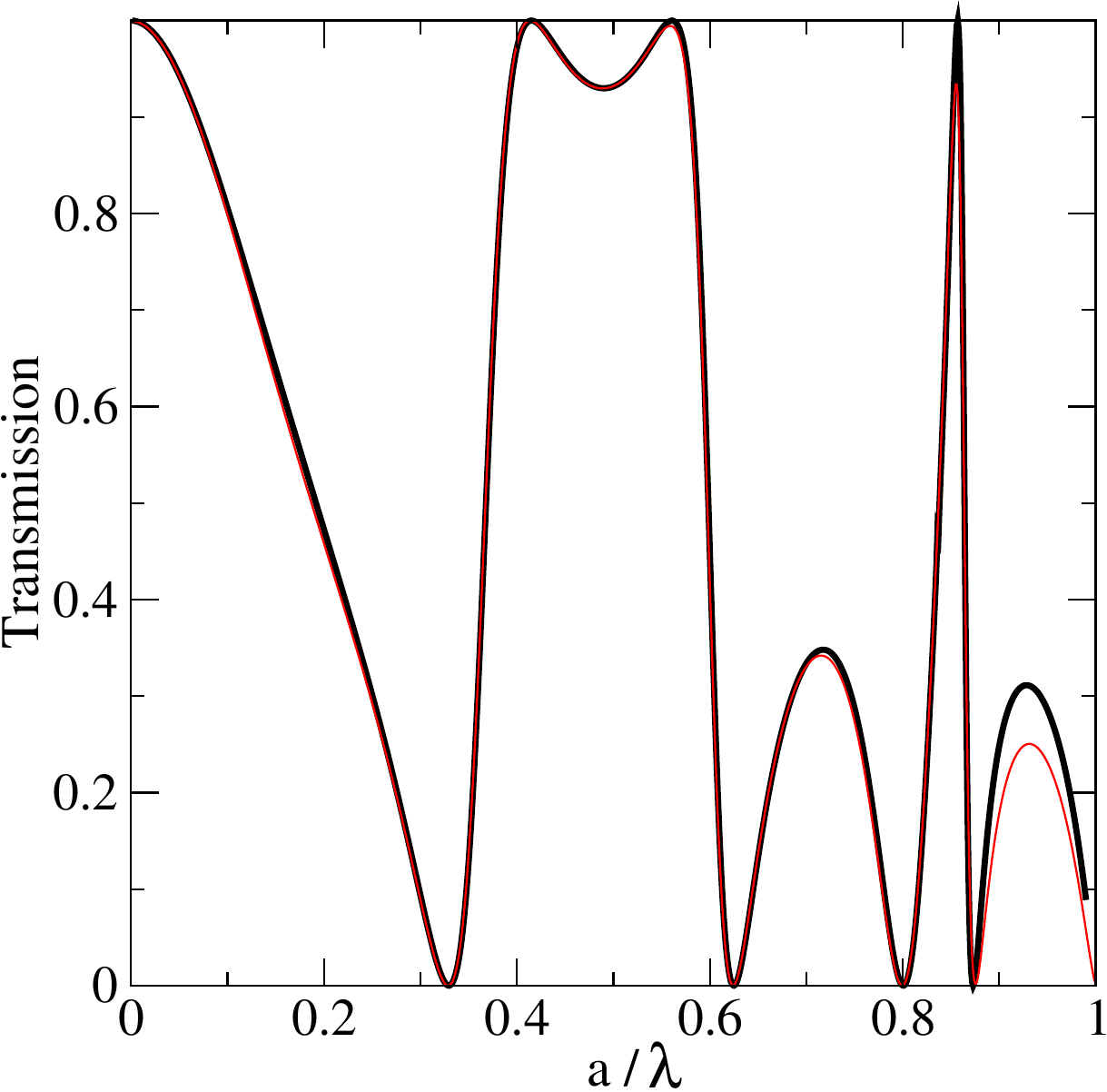}
~~~~~
\includegraphics[width=0.35\textwidth]{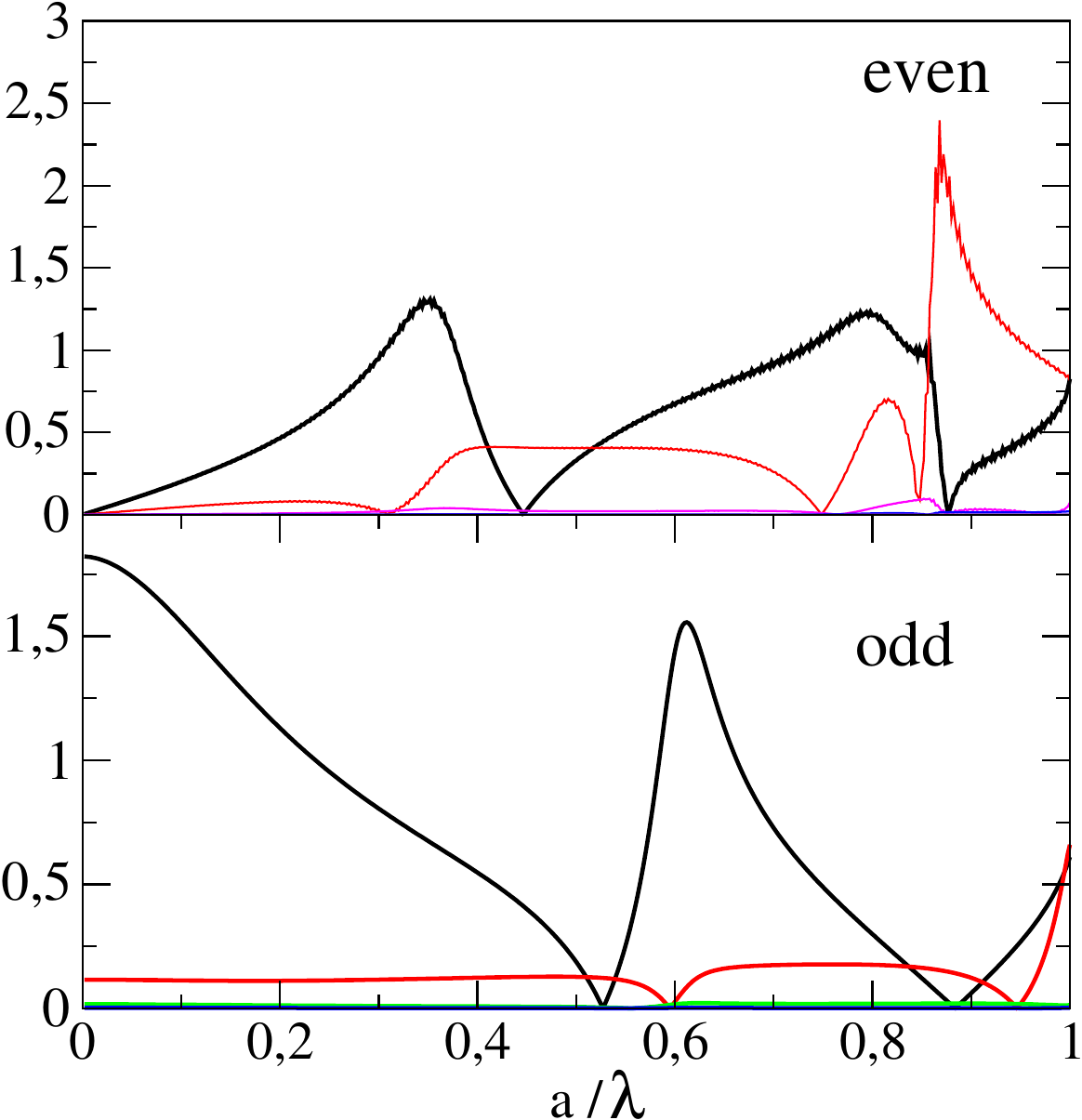}
\caption{Left: Transmission of $H_z$-polarized plane wave through the chain of left-handed cylinders
calculated by the transfer matrix method and from equation \ref{eq:T}. Similarly to the case of dielectric cylinders,
two numerical  methods give identical result, because Fano resonances (shown in the right panel)
are much weaker than in the case of $E_z$ polarized wave (Fig. \ref{fig:lhm-t}).
}
\end{center}
\label{fig:lhm-h}
\end{figure}

\subsection{Left-handed  cylinders}

The same analysis of the transmission spectra of cylinders made from the left-handed medium ($\varepsilon = -12$,
$\mu=-1$)
is more difficult, since standard numerical techniques (transfer matrix method, RCWA ) seem not to be
suitable for the
calculation of the transmission coefficient. As shown in the left panel of figure  \ref{fig:lhm-t}, results obtained by the transfer matrix method
with  three  discretization of the unit cell provide us with completely different
transmission coefficients.  Such failure, 
obtained also when the transmission was calculated by the RCWA method indicates that the eigenstates 
of the electromagnetic field inside  the structure are
strongly inhomogeneous.  Consequently, excited guided modes cannot be expanded in a finite series 
in the the basis of eigenfrequences, used in the transfer matrix 
method. 

The assumption of strong spatial inhomogeneity  is confirmed  by frequency dependence of coefficients $\beta$ shown in
middle panel of figure  \ref{fig:lhm-t}.  Incident electromagnetic wave
excites in the left-handed  chain a series of guided modes. The resonant frequencies of these modes 
lie very close to each other (in fact, since
the resonances have a finite width, they overlap). Also, absolute values of coefficients $\beta$ are in order of magnitude 
larger than in the case of dielectric cylinder.

Right panel of figure \ref{fig:lhm-t} shows the transmission coefficient  obtained by equation (\ref{eq:T}).
 Two minims in the transmission coefficient coincide  with   eigenfrequences of excitation of guided modes.   
The spatial distribution of electric field shown in the bottom panel confirm that many resonances are excited simultaneously
 -- in contrast to dielectric structure, we cannot easily identify the order of excited modes.

\medskip

The above mentioned problem of numerical instability is  not actual  when transmission of the  
$H_z$ polarized wave is calculated (figure 9). 
Now, numerical data obtained by the  transfer matrix method agree perfectly with those found by the present method 
(left panel).  As shown in the right panel, resonances of  guided modes are very weak and broad, so that they have negligible
influence on the transmission coefficient.

\section{Conclusion}\label{sect:concl}

We presented physical and numerical analysis of the guided modes of linear chain of cylinders, made either from
the dielectric or from the left-handed material. Comparison of these two structures shows 
new phenomena could be observed when permittivity and permeability possess negative values. We proved that the 
transmission of electromagnetic waves through the photonic structure with left-handed components is strongly 
influenced by guided modes excited in the structure.

The spectrum of guided modes  contains much more branches. In contrast to dielectric structure, the eigenfrequency of
these modes decreases when mode index $k$ increases. Also, there are no guided modes for the wave vector
close to some critical value $q_c$. We suppose  that this absence of guided modes
corresponds to the folding of frequency bands observed recently \cite{chen,busch}.

Very rich spectrum of guided modes is responsible for numerical instabilities when  the transmission coefficients
is calculated by standard methods.
Since eigenmodes of the left-handed structure strongly differs from plane waves, any numerical algorithm, based on the expansion
of the fields into the plane waves (transfer matrix or RCWA) fails to recover true transmission coefficient.

\section{Appendix A}
For  $k,m= 1,2,\dots N_B$,
$J_k \equiv J_k(2\pi R/\lambda)$
and ${\cal H}_k$  given by equation \ref{eq:haha}
we obtain 
the $N+1\times N+1$ matrix \textbf{B} 
\begin{equation}\label{eq:B}
\begin{array}{lcl}
B_{00} &=& {\calh}_0 + 
\displaystyle{\sum_{n=1}^{N_s}}~~
2\cos q H_0(un)J_0\\
B_{0m} &=& 
\displaystyle{\sum_{n=1}^{N_s}}~~
H_m(un)J_0 \times [(-1)^m e^{iqan}+ e^{-iqan} ]\\
B_{k0} &=& 
\displaystyle{\sum_{n=1}^{N_s}}~~
H_k(un)J_k \times 2[ e^{iqan} + (-1)^ke^{-iqan} ] \\
  && \\
B_{km} &=& 
2{\calh}_k\delta_{km} + 
\displaystyle{\sum_{n=1}^{N_s}}~~
[e^{iqan}(-1)^{m-k} + e^{-iqan}][H_{m-k}(un)+(-1)^{k}H_{m+k}(un)] J_k
\end{array}
\end{equation}
and the $N\times N$ matrix \textbf{C}:
\begin{equation}
\begin{array}{lcl}
\label{eq:C}
C_{km} = 2{\calh}_k\delta_{km} + 
\displaystyle{\sum_{n=1}^{N_s}}&&
[e^{iqan}(-1)^{m-k} + e^{-iqan}]\\
&\times& [{\calh}_{m-k}(un)-(-1)^{k}{\calh}_{m+k}(un)] J_k
\end{array}
\end{equation}
Matrices \textbf{B'} and \textbf{C'} could be obtained from \textbf{B} and \textbf{C}, respectively, by substitutions 
\begin{equation}
J_k \to J'_k, ~~~~~{\calh}_k\to{{\calh}'}_k
\end{equation}

\section*{Acknowledgment}

This work was supported by the Slovak Research and Development Agency under the contract No. APVV-0108-11
and by the Agency  VEGA under the contract No. 1/0372/13.

\end{document}